\documentstyle[12pt]{article}
\input{psfig.sty}

\def\IB{\relax\hbox{$\inbar\kern-.3em{\rm B}$}}
\def\IC{\relax\hbox{$\inbar\kern-.3em{\rm C}$}}
\def\ID{\relax\hbox{$\inbar\kern-.3em{\rm D}$}}
\def\IE{\relax\hbox{$\inbar\kern-.3em{\rm E}$}}
\def\IF{\relax\hbox{$\inbar\kern-.3em{\rm F}$}}
\def\IG{\relax\hbox{$\inbar\kern-.3em{\rm G}$}}
\def\IGa{\relax\hbox{${\rm I}\kern-.18em\Gamma$}}
\def\IH{\relax{\rm I\kern-.18em H}}
\def\IK{\relax{\rm I\kern-.18em K}}
\def\IL{\relax{\rm I\kern-.18em L}}
\def\IP{\relax{\rm I\kern-.18em P}}
\def\IR{\relax{\rm I\kern-.18em R}}
\def\IZ{\relax\ifmmode\mathchoice
{\hbox{\cmss Z\kern-.4em Z}}{\hbox{\cmss Z\kern-.4em Z}}
{\lower.9pt\hbox{\cmsss Z\kern-.4em Z}}
{\lower1.2pt\hbox{\cmsss Z\kern-.4em Z}}\else{\cmss Z\kern-.4em Z}\fi}
\def\II{\relax{\rm I\kern-.18em I}}

\def\sCC{{\kern 0.27em\vrule height1.45ex width0.03em depth0em
          \kern-0.30em\rm C}}
\def\C{{\mathchoice
  {\sCC}
  {\sCC}
  {\kern 0.225em \vrule height1.05ex width0.025em depth0em \kern-0.25em \rm C}
  {\kern 0.180em \vrule height0.78ex width0.02em depth0em \kern-0.2em \rm C}
        }}
\def\sHH{{\rm I\kern-.16em{}H}}
\def\H{{\mathchoice
  {\sHH}
  {\sHH}
  {\rm I\kern-.13em{}H}
  {\rm I\kern-.13em{}H} }}
\def\sNN{{\rm I\kern-.16em{}N}}
\def\N{{\mathchoice
  {\sNN}
  {\sNN}
  {\rm I\kern-.12em{}N}
  {\rm I\kern-.10em{}N} }}
\def\sPP{{\rm I\kern-.16em{}P}}
\def\P{{\mathchoice
  {\sPP}
  {\sPP}
  {\rm I\kern-.12em{}P}
  {\rm I\kern-.10em{}P} }}
\def\sQQ{{\kern 0.27em \vrule height1.45ex width0.03em depth0em
          \kern-0.30em \rm Q}}
\def\Q{{\mathchoice
        {\sQQ}
        {\sQQ}
  {\kern 0.225em \vrule height1.05ex width0.025em depth0em \kern-0.25em \rm Q}
  {\kern 0.180em \vrule height0.78ex width0.020em depth0em \kern-0.20em \rm Q}
        }}
\def\sRR{{\rm I\kern-0.16em{}R}}
\def\R{{\mathchoice
  {\sRR}
  {\sRR}
  {\rm I\kern-0.12em{}R}
  {\rm I\kern-0.10em{}R} }}
\def\sZZ{{\rm Z\kern-0.32em{}Z}}
\def\Z{{\mathchoice
  {\sZZ}
  {\sZZ} 
  {\rm Z\kern-0.3em{}Z}     
  {\rm Z\kern-0.25em{}Z} }}  
\def\ZZZ{{\rm Z\kern-0.24em{}Z}}

\def\Tr{{\rm Tr}}

\def\dim{{\rm dim}}

\def\inbar{\,\vrule height1.5ex width.4pt depth0pt}
\font\cmss=cmss10 \font\cmsss=cmss10 at 7pt

\begin{document}
\renewcommand{\baselinestretch}{1}

\thispagestyle{empty}
{\flushright{\small MIT-CTP-2803\\hep-th/9811183\\}}

\vspace{.3in}
\begin{center}\LARGE {Non-Abelian Finite Gauge Theories}
\end{center}

\vspace{.2in}
\begin{center}
{\large Amihay Hanany and Yang-Hui He\\}
\normalsize{hanany, yhe@ctp.mit.edu\\}
\vspace{.2in} {\it Center for Theoretical Physics,\\ Massachusetts
Institute of Technology\\ Cambridge, Massachusetts 02139, U.S.A.\\}
\vspace{.2in} November, 1998\\
\end{center}

\vspace{0.1in}
\begin{abstract}
We study orbifolds of ${\cal N} = 4$ $U(n)$ super-Yang-Mills theory
given by discrete subgroups of $SU(2)$ and $SU(3)$.
We have reached many interesting observations that have
graph-theoretic interpretations. For the subgroups of $SU(2)$, we have 
shown how the matter content agrees with current quiver theories
and have offered a possible explanation. In the case of $SU(3)$ we have
constructed
a catalogue of candidates for finite (chiral) ${\cal N}=1$ theories, giving
the gauge group and matter content. Finally, we
conjecture a McKay-type correspondence for Gorenstein singularities in
dimension 3 with modular invariants of WZW conformal models.
This implies a connection between a class of finite ${\cal N}=1$ supersymmetric
gauge theories in four dimensions and
the classification of affine $SU(3)$ modular invariant partition functions
in two dimensions.
\end{abstract}

\section{Introduction}
Recent advances on finite four dimensional gauge theories from string theory
constructions have been dichotomous: either from the geometrical perspective of
studying algebro-geometric singularities such as orbifolds
\cite{KS} \cite{LNV} \cite{BKV}, 
or from the intuitive perspective of studying various configurations of branes 
such as the so-called brane-box models \cite{HZ}.
(See \cite{HU} and references therein for a detailed description of these
models. A recent paper discusses the bending of non-finite
models in this context \cite{LR}.)
The two approaches lead to the realisation of finite, possibly chiral, ${\cal N}=1$
supersymmetric gauge theories, such as those discussed in
\cite{Leigh}.
Our ultimate dream is of course to have the flexbility of the equivalence and
completion of these approaches, allowing us to compute say, the duality group
acting on the moduli space of marginal gauge couplings \cite{HSU}.
(The duality groups for the ${\cal N}=2$ supersymmetric theories were
discussed in the context of these two approaches in \cite{KMV} and 
\cite{Witten1}.)
The brane-box method has met great success in providing the intuitive picture
for orbifolds by Abelian groups: the elliptic model consisting of 
$k \times k'$ branes conveniently reproduces the theories on
orbifolds by $\IZ_k \times \IZ_{k'}$ \cite{HU}.
Orbifolds by $\IZ_k$ subgroups of $SU(3)$ are given by Brane Box Models with
non-trivial identification on the torus \cite{HSU} \cite{HU}.
Since by the structure
theorem that all finite Abelian groups are direct sums of cyclic ones, 
this procedure
can be presumably extended to all Abelian quotient singularities.
The non-Abelian groups however, present difficulties. By adding orientifold planes,
the dihedral groups have also been successfully attacked for theories with
${\cal N}=2$ supersymmetry \cite{Kapustin}. The
question still remains as to what could be done for the myriad of finite groups,
and thus to general Gorenstein singularities.

In this paper we shall present a catalogue of these Gorenstein
singularities in dimensions 2 and 3, i.e., orbifolds constructed from 
discrete subgroups of $SU(2)$ and $SU(3)$ whose classification are complete.
In particular we shall concentrate on the gauge group, the fermionic and
bosonic matter content resulting from the orbifolding of an ${\cal N} = 4$
$U(n)$ super-Yang-Mills theory. In Section 2, we present the general arguments
that dictate the matter content for arbitrary finite group $\Gamma$. Then in
Section 3, we study the case of $\Gamma \subset SU(2)$ where we notice
interesting graph-theoretic descriptions of the matter matrices. 
We analogously analyse case by case, the discrete subgroups of $SU(3)$ in
Section 4, followed by a brief digression of possible mathematical 
interest in Section 5. This leads to a Mckay-type
connection between the classification
of two dimensional $SU(3)_k$ modular invariant partition functions and the
class of finite ${\cal N}=1$ supersymmetric gauge theories calculated in this
paper. 
Finally we tabulate possible chiral theories
obtainable by such orbifolding techniques for these $SU(3)$ subgroups.

\section {The Orbifolding Technique}
Prompted by works by Douglas, Greene, Moore and Morrison
on gauge theories which arise by placing D3 branes on orbifold singularities
\cite{Douglas} \cite{DG}, \cite{DGM},
Kachru and Silverstein \cite{KS} and subsequently
Lawrence, Nekrasov and Vafa \cite{LNV} noted that an orbifold theory 
involving the
projection of a supersymmetric ${\cal N}=4$ gauge theory on some discrete subgroup 
$\Gamma \subset SU(4)$ leads to a conformal field theory with 
${\cal N}\leq 4$ supersymmetry. 
We shall first briefly summarise their results here.

We begin with a $U(n)$ ${\cal N}=4$ super-Yang-Mills
theory which has an R-symmetry
of $Spin(6)\simeq SU(4)$. There are gauge bosons $A_{IJ}$ $(I,J=1,...,n)$
being singlets of $Spin(6)$, along with
adjoint Weyl fermions $\Psi _{IJ}^{\bf{4}}$
in the fundamental $\bf{4}$ of $SU(4)$ and adjoint scalars $\Phi _{IJ}^{%
\bf{6}}$ in the antisymmetric $\bf{6}$ of $SU(4)$. Then we choose a
discrete (finite) subgroup $\Gamma \subset SU(4)$ with the set of
irreducible representations $\left\{ {\bf r}_{i}\right\}$ acting on
the gauge group by breaking the $I$-indices up according to 
$\left\{ {\bf r}_{i}\right\} $, i.e., by 
$\bigoplus\limits_{i} {\bf r}_{i}=\bigoplus\limits_{i}%
 \IC^{N_{i}} {\bf r}_{i}$ such that $\IC^{N_{i}}$ 
accounts for the multiplicity of each ${\bf r}_{i}$ and 
$n=\sum\limits_{i=1}N_{i}\dim ({\bf r}_{i})$.
In the string theory picture, this decomposition of the gauge group
corresponds to permuting $n$ D3-branes and hence their Chan-Paton factors
which contain the $IJ$ indices,
on orbifolds of $\IR^{6}$. Subsequently by the Maldecena large $N$ conjecture
\cite{Maldecena}, we have
an orbifold theory on $AdS_{5}\times S^{5}$, with the R-symmetry
manifesting as the $SO(6)$ symmetry group of $S^{5}$ in which the 
branes now live \cite{KS}. The string perturbative calculation 
in this context, especially with respect to vanishing theorems for $\beta$-functions,
has been performed \cite{BKV}.

Having decomposed the gauge group, we must likewise do so for the matter 
fields: since an orbifold is invariant under the $\Gamma$-action,
we perform the so-called projection on the fields by keeping only the 
$\Gamma $-invariant fields in the theory. Subsequently we arrive at a
(superconformal) field theory with gauge group 
$G=\bigotimes\limits_{i}SU(N_{i})$ and Yukawa and quartic interaction
respectively as (in the notation of \cite{LNV}):
\[
\begin{array}{l}
Y = \sum_{ij k} \gamma_{ijk}^{f_{ij}, f_{jk}, f_{ki}} \Tr
\Psi_{f_{ij}}^{ij} \Phi_{f_{jk}}^{jk} \Psi_{f_{ki}}^{ki} \\ \\

V = \sum_{ijkl} \eta^{ijkl}_{f_{ij}, f_{jk}, f_{kl},
f_{li}} \Tr
\Phi_{f_{ij}}^{ij}\Phi_{f_{jk}}^{jk}\Phi_{f_{kl}}^{kl}\Phi_{f_{li}}^{li},
\end{array}
\]
where 
\[
\begin{array}{l}
\gamma_{ij k}^{f_{ij}, f_{jk}, f_{ki}} =
\Gamma_{\alpha\beta, m} \left( Y_{f_{ij}}
\right)^{\alpha}_{v_{i}{\bar v}_{j}}
\left( Y_{f_{jk}} \right)^{m}_{v_{j}{\bar v}_{k}} \left( Y_{f_{ki}}
\right)^{\beta}_{v_{k}{\bar v}_{i}} \\ \\

\eta^{ijkl}_{f_{ij}, f_{jk}, f_{kl}, f_{li}} =
\left(Y_{f_{ij}}\right)^{[ m}_{v_{i}{\bar v}_{j}}
\left(Y_{f_{jk}}\right)^{n]}_{v_{j}{\bar v}_{k}}
\left(Y_{f_{kl}}\right)^{[m}_{v_{k}{\bar v}_{l}}
\left(Y_{f_{li}}\right)^{n]}_{v_{l}{\bar v}_{i}},
\end{array}
\] 
such that $\left( Y_{f_{ij}} \right)^{\alpha}_{v_{i}{\bar v}_{j}}$,
$\left(Y_{f_{ij}}\right)^{m}_{v_{i}{\bar v}_{j}}$ are the $f_{ij}$'th
Clebsch-Gordan
coefficients corresponding to the projection of $4\otimes {\bf r}_i$
and $6\otimes {\bf r}_i$ onto
${\bf r}_j$, and $\Gamma_{\alpha\beta, m}$ is the invariant in
${\bf 4} \otimes {\bf 4} \otimes {\bf 6}$.

Furthermore, the matter content is as follows:
\begin{enumerate}
\item  Gauge bosons transforming as
\[
\hom \left( \IC^{n},\IC^{n}\right) ^{\Gamma}
=\bigoplus\limits_{i}\IC^{N_{i}} {\bf \otimes }\left( \IC^{N_{i}}\right) ^{*},
\]
which simply means that the original (R-singlet) adjoint $U(n)$ fields now
break up according to the action of $\Gamma$ to become the 
adjoints of the various $SU(N_i)$;

\item  $a_{ij}^{\bf{4}}$ Weyl fermions
 $\Psi _{f_{ij}}^{ij}$
($f_{ij}=1,...,a_{ij}^{\bf{4}}$ ) 
\[
\left( {\bf 4} \otimes \hom \left(\IC^{n},\IC^{n}\right) \right) ^{\Gamma
}=\bigoplus\limits_{ij}a_{ij}^{\bf 4}\IC^{N_{i}} {\bf \otimes }
\left( \IC^{N_{j}}\right) ^{*},
\]
which means that these fermions in the fundamental {\bf 4} of the 
original R-symmetry now become $\left(N_{i},\overline{N}_{j}\right)$
bi-fundamentals of $G$ and there are $a_{ij}^{\bf 4}$ copies of them;

\item  $a_{ij}^{\bf 6}$ scalars $\Phi _{f_{ij}}^{ij}$
($f_{ij}=1,...,a_{ij}^{\bf{6}}$ ) as
\[
\left( {\bf 6} \otimes \hom \left(\IC^{n},\IC^{n}\right) \right) ^{\Gamma}
= \bigoplus\limits_{ij}a_{ij}^{\bf{6}}\IC^{N_{i}} {\bf \otimes }
\left( \IC^{N_{j}}\right)^{*},
\]
similarly, these are $G$ bi-fundamental bosons, inherited from the
{\bf 6} of the original R-symmetry.
\end{enumerate}

For the above, we define $a_{ij}^{\cal R}$ (${\cal R} = {\bf 4}$ or {\bf 6}
for fermions and bosons respectively) as the composition coefficients
\begin{equation}
{\cal R}\otimes {\bf r}_{i}=\bigoplus\limits_{j}a_{ij}^{{\cal R}}
{\bf r}_{j}
\label{aij}
\end{equation}

Moreover, the supersymmetry of the projected theory must have its R-symmetry
in the commutant of $\Gamma \subset SU(4)$, which is $U(2)$ for 
$SU(2)$, $U(1)$ for $SU(3)$ and trivial for $SU(4)$, which means:
if $\Gamma \subset SU(2)$, we have an ${\cal N} = 2$ theory, 
if $\Gamma \subset SU(3)$, 
we have ${\cal N} = 1$, and finally for $\Gamma \subset $ the full $SU(4)$, 
we have a non-supersymmetric theory.

Taking the character $\chi $ for element $\gamma \in \Gamma $ on both sides
of (\ref{aij}) and recalling that $\chi $ is a $\left( \otimes
,\oplus \right) $-ring homomorphism, we have
\begin{equation}
\chi _{\gamma }^{{\cal R}}\chi _{\gamma
}^{(i)}=\sum\limits_{j=1}^{r}a_{ij}^{{\cal R}}\chi _{\gamma }^{(j)}
\label{aijchar}
\end{equation}
where $r=\left| \left\{ {\bf r}_{i}\right\} \right| $, the number of
irreducible representations, which by an elementary theorem on finite
characters, is equal to the number of inequivalent conjugacy classes of
$\Gamma$. We further recall the orthogonality theorem of finite characters,

\begin{equation}
\sum\limits_{\gamma =1}^{r}r_{\gamma }\chi_{\gamma }^{(i)*}
	\chi_{\gamma }^{(j)}=g\delta^{ij},
\label{ortho}
\end{equation}

where $g=\left| \Gamma \right| $ is the
order of the group and  $r_{\gamma }$ is the order of the conjugacy class
containing $\gamma$. Indeed, $\chi$ is a class function and is hence constant
for each conjugacy class; moreover, 
$\sum\limits_{\gamma =1}^{r}r_{\gamma }=g$ is the class equation for
$\Gamma$. This orthogonality
allows us to invert (\ref{aijchar}) to finally give the matrix $a_{ij}$ 
for the matter content
\begin{equation}
a_{ij}^{{\cal R}}=\frac{1}{g}\sum\limits_{\gamma
=1}^{r}r_{\gamma }\chi_{\gamma }^{{\cal R}}\chi_{\gamma }^{(i)}\chi
_{\gamma }^{(j)*}  \label{matter}
\end{equation}
where ${\cal R}$ $=\bf{4}\ $for Weyl fermions and  $\bf{6}$ for
adjoint scalars and the sum is effectively that over the columns of the
Character Table of $\Gamma $. Thus equipped, let us specialise to $\Gamma$
being finite discrete subgroups of $SU(2)$ and $SU((3)$.

\section{Checks for $SU(2)$}
The subgroups of $SU(2)$ have long been classified \cite{Klein};
discussions and applications thereof can be found in \cite{Mckay}
\cite{Seiberg} \cite{Kronheimer} \cite{Blichfeldt}. 
To algebraic geometers they give rise to 
the so-called Klein singularities and are labeled by the first affine
extension of the simply-laced simple Lie groups $\widehat{A}\widehat{D}
\widehat{E}$ (whose associated Dynkin diagrams are those of $ADE$ adjointed
by an extra node), i.e., there are two infinite series and 3 exceptional
cases:

\begin{enumerate}
\item  $\widehat{A}_{n}=\IZ_{n+1}$, the cyclic group of order $n+1$;

\item  $\widehat{D}_{n}$, the binary lift of the ordinary dihedral group $d_n$;

\item  the three exceptional cases, $\widehat{E}_{6}$, $\widehat{E}_{7}$ and 
$\widehat{E}_{8}$, the so-called binary or double\footnote{
For $SO(3)\cong SU(2)/\IZ_2$ these would be the familiar 
symmetry groups of the respective
regular solids in $\IR^{3}$: the dihedron, tetrahedron, octahedron/cube and
icosahedron/dodecahedron. 
However since we are in the double cover
$SU(2),$ there is a non-trivial $\IZ_{2}$- lifting,\\
$
\begin{array}{ccccccccc}
0 & \rightarrow & \IZ_2 & \rightarrow & SU(2) & \rightarrow & SO(3) & \rightarrow
& 0, \\
  & & & & \bigcup  & & \bigcup  & &\\
  & & & & {\widehat D}, {\cal T}, {\cal O}, {\cal I} & \rightarrow &d, T, O, I & & \\
\end{array}
$ \\
hence the modifier ``binary''. Of course, the $A$-series, being abelian, receives no
lifting. Later on we shall briefly touch upon the
ordinary $d,T,I,O$ groups as well.} tetrahedral, 
octahedral and icosahedral groups ${\cal T, O, I}$.
\end{enumerate}

The character tables for these groups are known \cite{Lomont} \cite{Zaf} \cite{CFT} and
are included in Appendix I for reference.
Therefore to obtain (\ref{matter}) the only difficulty remains in the
choice of ${\cal R}$. We know that whatever ${\cal R}$ is, it must be 4 
dimensional for the fermions and 6 dimensional for the bosons inherited from
the fundamental {\bf 4} and antisymmetric {\bf 6} of $SU(4)$.
Such an ${\cal R}$ must therefore be a 4 (or 6) dimensional irrep of 
$\Gamma$, or be the tensor sum of lower dimensional irreps (and hence be
reducible); for the character table, this means that the row
of characters for ${\cal R}$ (extending over the conjugacy 
classes of $\Gamma$) must be an existing row or the sum of
existing rows.
Now since the first column of the character table of any finite group
precisely gives
the dimension of the corresponding representation, it must therefore be that
$\dim({\cal R}) = 4,6$ should be partitioned into these numbers.
Out of these possibilities we must select the one(s)
consistent with the decomposition of the {\bf 4} and {\bf 6} of $SU(4)$ into the
$SU(2)$ subgroup\footnote{We note that even though 
this decomposition is that into irreducibles
for the full continuous Lie groups, such irreducibility may not be inherited by 
the discrete subgroup, i.e., the {\bf 2}'s {\it may} not be irreducible representations
of the finite $\Gamma$.}, namely:

\begin{equation}
\begin{array}{ccc}
SU(4) & \rightarrow & SU(2) \times SU(2) \times U(1) \\
{\bf 4} & \rightarrow & ({\bf 2,1})_{+1} \bigoplus ({\bf 1,2})_{-1} \\
{\bf 6} & \rightarrow & ({\bf 1,1})_{+2} \bigoplus ({\bf 1,1})_{-2} \bigoplus 
			({\bf 2,2})_0 \\
\end{array}
\label{break2}
\end{equation}

where the subscripts correspond to the $U(1)$ factors (i.e., the trace) and in
particular the $\pm$ forces the overall traceless condition.
 From (\ref{break2}) we know that $\Gamma \subset SU(2)$ inherits a {\bf 2}
while the complement is trivial. This means that the 4 dimensional represention
of $\Gamma$ must be decomposable into a nontrivial 2 dimensional one with a 
trivial 2 dimensional one. In the character language, this means that
${\cal R} = {\bf 4 = 2_{trivial} \oplus 2 } $ where
${\bf 2_{trivial} = 1_{trivial} \oplus 1_{trivial}}$, the tensor sum of two copies
of the (trivial) principal representation where all group elements are mapped to the
identity, i.e., corresponding to the first row in the character table. Whereas for the 
bosonic case we have
${\cal R} = {\bf 6 = 2_{trivial}
 \oplus 2 \oplus 2^{'} } $. We have denoted ${\bf 2^{'}}$ to signify that the
two {\bf 2}'s may not be the same, and correspond to inequivalent 
representations of $\Gamma$ with the same dimension. However we can restrict this 
further by recalling that the antisymmetrised tensor product 
$[{\bf 4} \otimes {\bf 4}]_A \rightarrow
1 \oplus 2  \oplus 2 \oplus [2 \otimes 2]_A$ must
in fact contain the {\bf 6}. Whence we conclude that ${\bf 2} = {\bf 2^{'}}$.
Now let us again exploit the additive property of the group character, i.e., a
homomorphism from a $\oplus$-ring to a +-subring of a number field (and indeed
much work has been done for the subgroups in the case of number fields of various
characteristics); this
means that we can simplify $\chi^{{\cal R}=x \oplus y}$ as $\chi^x + \chi^y$.
Consequently, our matter matrices become:
\[
\begin{array}{l}
a_{ij}^{{\bf 4}}=\frac{1}{g}\sum\limits_{\gamma=1}^{r}r_{\gamma} 
	\left(2 \chi_{\gamma}^{{\bf 1}} + \chi_{\gamma}^{{\bf 2}} \right)
	\chi_{\gamma }^{(i)}\chi_{\gamma }^{(j)*}
	= 2 \delta_{ij} + \frac{1}{g}\sum\limits_{\gamma=1}^{r}r_{\gamma} 
	\chi_{\gamma}^{{\bf 2}}
	\chi_{\gamma }^{(i)}\chi_{\gamma }^{(j)*}\\

a_{ij}^{{\bf 6}}=\frac{1}{g}\sum\limits_{\gamma=1}^{r}r_{\gamma} 
	\left(2 \chi_{\gamma}^{{\bf 1}} + \chi_{\gamma}^{{\bf 2}
	\oplus {\bf 2}} \right)
	\chi_{\gamma }^{(i)}\chi_{\gamma }^{(j)*}
	= 2 \delta_{ij} + \frac{2}{g}\sum\limits_{\gamma=1}^{r}r_{\gamma} 
	\chi_{\gamma}^{{\bf 2}} \chi_{\gamma }^{(i)}\chi_{\gamma }^{(j)*}\\
\end{array}
\label{matter2}
\]
where we have used the fact that $\chi$ of the trivial representation are
all equal to 1, thus giving by (\ref{ortho}), the $\delta_{ij}$'s. 
This simplification thus limits our attention to only 2 dimensional 
representations of $\Gamma$; however there still may remain many possibilities
since the {\bf 2} may be decomposed into nontrivial {\bf 1}'s or there may
exist many inequivalent irreducible {\bf 2}'s.

We now appeal to physics for further restriction. We know that the ${\cal N}
= 2$ theory (which we recall is the resulting case when $\Gamma \subset 
SU(2)$) is a non-chiral supersymmetric theory; this means our bifundamental
fields should not distinguish the left and right indices, i.e., the matter
matrix $a_{ij}$ must be {\it symmetric}.
Also we know that in the ${\cal N}=2$ vector multiplet there are 2 Weyl
fermions and 2 real scalars, thus the fermionic and bosonic matter matrices
have the same entries on the diagonal. Furthermore the hypermultiplet
has 2 scalars and 1 Weyl fermion in $(N_i, \bar{N}_j)$ and another 2 
scalars and 1 Weyl fermion in the complex conjugate representation, whence
we can restrict the off-diagonals as well, viz., $2 a_{ij}^{\bf 4} -
a_{ij}^{\bf 6}$ must be some multiple of the identity.
This supersymmetry
matching is of course consistent with (\ref{matter2}).

Enough said on generalities. Let us analyse the groups case by case.
For the cyclic group, the {\bf 2} must come from the tensor sum of two {\bf 1}'s.
Of all the possibilities, only the pairing of dual representations gives
symmetric $a_{ij}$. By dual we mean the two {\bf 1}'s which are
complex conjugates of each other (this of course includes when
${\bf 2} = {\bf 1_{trivial}^2}$,
which exist for all groups and gives us merely $\delta_{ij}$'s and can
henceforth be eliminated as uninteresting). We denote the nontrivial pairs
as ${\bf 1^{'}}$ and ${\bf 1^{''}}$. In this case we can easily perform
yet another consistency check. From (\ref{break2}), we have a traceless
condition seen as the cancelation of the $U(1)$ factors. That was on the
Lie algebra level; as groups, this is our familiar determinant unity 
condition. 
Since in the block decomposition (\ref{break2}) the 
${\bf 2_{trivial}} \subset$ the complement $SU(4)\backslash\Gamma$ 
clearly has determinant 1, this forces our {\bf 2} matrix
to have determinant 1 as well. However in this cyclic case, $\Gamma$ is
abelian, whence the characters are simply presentations of the group,
making the {\bf 2} to be in fact diagonal. Thus the determinant is simply the
product of the entries of the two rows in the character table. And indeed
we see for dual representations, being complex conjugate roots of unity,
the two rows do multiply to 1 for all members. Furthermore we note that
different dual pairs give $a_{ij}$'s that are mere permutations of each other.
We conclude that the fermion
matrix arises from ${\bf 1^2} \oplus {\bf 1^{'}} \oplus {\bf 1^{''}}$. For
the bosonic matrix, by (\ref{matter2}), we have  ${\bf 6} =
({\bf 1} \oplus {\bf 1^{'}} \oplus {\bf 1^{''}})^{\bf 2}$. These and ensuing
$a_{ij}$'s are included in Appendix II.

For the dihedral case, the {\bf 1}'s are all dual to the principal, corresponding
to some $\IZ_2$ inner automorphism among the conjugacy classes and the
characters consist no more than $\pm 1$'s, giving us $a_{ij}$'s which are
block diagonal in $((1,0),(0,1))$ or $((0,1),(1,0))$ and are not terribly 
interesting. Let us rigorise this statement. Whenever we have the character
table consisting of a row that is composed of cycles of roots of unity,
which is a persistent theme for {\bf 1} irreps, this corresponds in general to
some $\IZ_k$ action on the conjugacy classes. This implies that our
$a_{ij}$ for this choice of {\bf 1} will be the Kronecker product of matrices
obtained from the cyclic groups which offer us nothing new. We shall refer to 
these cases as ``blocks''; they offer us another condition of
elimination whose virtues we shall exploit much. 
In light of this, for the dihedral the choice of the {\bf 2} comes 
from the irreducible {\bf 2}'s which again give symmetric $a_{ij}$'s that
are permutations among themselves.
Hence ${\cal R} = {\bf 4} = {\bf 1^2} \oplus {\bf 2}$ and
${\cal R} = {\bf 6} = {\bf 1^2} \oplus {\bf 2^2}$.
For reference we have done likewise for the dihedral series not in the full 
$SU(2)$, the choice for ${\cal R}$ is the same for them.

Finally for the exceptionals ${\cal T,O,I}$, the {\bf 1}'s again give 
uninteresting block diagonals and out choice of {\bf 2} is again unique
up to permutation. Whence still 
${\cal R} = {\bf 4} = {\bf 1^2} \oplus {\bf 2}$ and
${\cal R} = {\bf 6} = {\bf 1^2} \oplus {\bf 2^2}$.
For reference we have computed the ordinary exceptionals $T,O,I$ which live in
$SU(2)$ with its center removed, i.e., in $SU(2)/\IZ_2 \cong SO(3)$. For them
the {\bf 2} comes from the ${\bf 1^{'}} \oplus {\bf 1^{''}}$, the {\bf 2},
and the trivial ${\bf 1^2}$ respectively.

Of course we can perform an {\it a posteriori} check. In 
this case of $SU(2)$ we already know the matter content due
to the works on quiver diagrams \cite{Douglas} \cite{Seiberg} \cite{Kapustin}.
The theory dictates that the 
matter content $a_{ij}$ can be obtained by looking at the Dynkin diagram of the
$\widehat{A} \widehat{D} \widehat{E}$ group associated to $\Gamma $ whereby 
one assigns 2 for $a_{ij}$ on the diagonal as well as 1 for
every pair of connected nodes
$ i \rightarrow j$ and $ 0 $ otherwise, i.e., $a_{ij}$ is essentially the
 adjacency matrix for the Dynkin diagrams treated as unoriented graphs.
Of course adjacency matrices for unoriented graphs are symmetric; this is
consistent with our nonchiral supersymmetry argument. Furthermore,
the dimension of $a_{ij}^{\bf 4}$ is required to be equal to the number of nodes 
in the associated affine Dynkin diagram (i.e., the rank).
This property is immediately seen to be satisfied by examining the character
tables in Appendix I where we note 
that the number of conjugacy classes of the respective finite groups 
(which we recall is equal to the number of irreducible representations) and
hence the dimension of $a_{ij}$ is indeed
that for the ranks of the associated affine algebras, namely $n+1$ for
$\widehat{A_n}$ and $\widehat{D_n}$ and 7,8,9 for $\widehat{E_{6,7,8}}$
respectively.
We note in passing that the conformality condition $N_f = 2 N_c$ for 
this ${\cal N}=2$ \cite{KS} \cite{LNV} nicely translates to the graph language:
it demands that for the one loop $\beta$-function to vanish the label of each
node
(the gauge fields) must be $\frac12$ that of those connected thereto (the
bi-fundamentals).

Our results for $a_{ij}$ computed using (\ref {matter}),
Appendix I, and the aforementioned decomposition of ${\cal R}$ are tabulated
in Appendix II. They are 
precisely in accordance with the quiver theory and present themselves as the
relevant adjacency matrices. One interesting point to note is that for the
dihedral series, the ordinary $d_n$ (which are in $SO(3)$ and not $SU(2)$)
for even $n$ also gave the binary $\widehat{D_{n'=\frac{n+6}{2}}}$ 
Dynkin diagram while the odd $n$ case always gave the ordinary 
$D_{n'=\frac{n+3}{2}}$ diagram.

These results should be of no surprise to us, since a similar calculation was
in fact done by J. Mckay when he first noted his famous correspondence \cite
{Mckay}. In the paper he computed the composition coefficients $m_{ij}$ in
$R \bigotimes R_j = \bigoplus\limits_{k} m_{jk} R_k$ for $\Gamma \subset
SU(2)$ with $R$ being a faithful representation thereof. He further noted that
for all these $\Gamma$'s there exists (unique up to automorphism) 
such $R$, which is precisely the 2 
dimensional irreducible representation for $\widehat{D}$ and $\widehat{E}$ 
whereas for $\widehat{A}$ it is the
direct sum of a pair of dual 1 dimensional representations. Indeed this is
exactly the decomposition of ${\cal R}$ which we have argued above from
supersymmetry. His {\it Theorema Egregium} was then\\

\noindent
{\large {\bf Theorem:}} The matrix $m_{ij}$ is $2 I$ minus
the cartan matrix, and is thus the adjacency matrix for the 
associated affine
Dynkin diagram treated as undirected $C_2$-graphs (i.e., maximal eigenvalue is 
2).\\

Whence $m_{ij}$ has 0 on the diagonal and 1 for connected nodes.
Now we note from our discussions above and results in Appendix II, 
that our $\cal R$ is precisely 
Mckay's $R$ (which we henceforth denote as $R_M$) plus two copies of the 
trivial representation for the {\bf 4} and
$R_M$ plus the two dimensional irreps in addition to the two copies of the 
trivial for the {\bf 6}. Therefore we conclude from (\ref{matter}):
\[
\begin{array}{l}
a_{ij}^{{\bf 4}}= \frac{1}{g}\sum\limits_{\gamma
	=1}^{r}r_{\gamma }\chi_{\gamma }^{R_M \oplus {\bf 1}^2 }
	\chi_{\gamma }^{(i)}
	\chi_{\gamma }^{(j)*} \\
a_{ij}^{{\bf 6}}=\frac{1}{g}\sum\limits_{\gamma
	=1}^{r}r_{\gamma }\chi_{\gamma }^{R_M \oplus R_M \oplus {\bf 1}^2 }
	\chi_{\gamma }^{(i)}\chi_{\gamma }^{(j)*} \\
\end{array}
\]
which implies of course, that our matter matrices should be
\[
\begin{array}{l}
a_{ij}^{{\bf 4}}= 2 \delta_{ij} + m_{ij} \\
a_{ij}^{{\bf 6}}= 2 \delta_{ij} + 2 m_{ij} \\
\end{array}
\]
with Mckay's $m_{ij}$ matrices. This is exactly the results
we have in Appendix II.  Having obtained such an elegant graph-theoretic
interpretation to our results, we remark that from this
point of view, oriented graphs means chiral gauge theory and connected means
interacting gauge theory. Hence we have the foresight that the ${\cal N} = 1$
case which we shall explore next will involve oriented graphs.

Now Mckay's theorem explains why the discrete subgroups of $SU(2)$ and hence
Klein singularities of algebraic surfaces (which our orbifolds essentially
are) as well as subsequent gauge theories thereupon afford this correspondence
with the affine simply-laced Lie groups. However they were originally proven
on a case by case basis, and we would like to know a deeper connection,
especially in light of quiver theories. We can partially answer this 
question by noting a beautiful
theorem due to Gabriel \cite{Gabriel} \cite{Bernstein} which forces the quiver
considerations by Douglas et al.\ \cite{Douglas}
to have the ADE results of Mckay.

It turns out to be convenient to formulate the theory axiomatically.
We define ${\cal L}(\gamma,\Lambda)$,
for a finite connected graph $\gamma$ with orientation $\Lambda$, vertices
$\gamma_0$ and edges $\gamma_1$, 
to be the category of quivers whose objects are any
collection $(V,f)$ of spaces $V_{\alpha \in \gamma_0}$ and mappings 
$f_{l \in \gamma_1}$ and whose morphisms are 
$\phi : (V,f) \rightarrow (V',f') $
a collection of linear mappings $\phi_{\alpha \in \Gamma_0} : V_{\alpha}
\rightarrow V'_{\alpha}$ compatible with $f$ by 
$\phi_{e(l)}f_l = f'_{l}\phi_{b(l)}$ where $b(l)$ and $e(l)$ are the beginning
and end of the directed edge $l$. Then we have\\

\noindent
{\large {\bf Theorem:}} If in the quiver category ${\cal L}(\gamma,\Lambda)$ 
there are only finitely many
non-isomorphic indecomposable objects, then $\gamma$ coincides with one of the
graphs $A_n,D_n,E_{6,7,8}$.\\

This theorem essentially compels any finite quiver theory to be constructible
only on graphs which are of the type of the Dynkin diagrams of $ADE$. And
indeed, the theories of Douglas, Moore et al.\ \cite{Douglas} \cite{Kapustin}
have explicitly made the physical realisations of these constructions.
We therefore see how Mckay's calculations, quiver theory and 
our present calculations nicely fit together for the case of $\Gamma
\subset SU(2) $.

\section{The case for $SU(3)$}
We repeat the above analysis for $\Gamma = SU(3)$, though now we have no 
quiver-type theories to aid us. The discrete subgroups of $SU(3)$ have also 
been long classified \cite{Blichfeldt}. They include (the order of these groups are
given by the subscript), other than all those of $SU(2)$ since $SU(2) \subset SU(3)$, 
the following new cases. We point out that in addition to the cyclic group in $SU(2)$, 
there is now in fact another Abelian case $\IZ_k \times \IZ_{k'}$ for $SU(3)$ 
generated by the matrix $((e^{\frac{2 \pi i}{k}},0,0),(0,e^{\frac{2 \pi i}{k'}},0),
(0,0,e^{-\frac{2 \pi i}{k}-\frac{2 \pi i}{k'}}))$ much in the spirit that
$((e^{\frac{2 \pi i}{n}},0),(0,e^{-\frac{2 \pi i}{n}}))$ generates the $\IZ_n$ for
$SU(2)$. Much work has been done for this $\IZ_k \times \IZ_{k'}$ case, q.\ v.\, 
\cite{HU} and references therein.
\begin{enumerate}
\item Two infinite series $\Delta_{3n^2}$ and $\Delta_{6n^2}$ for $n \in \IZ$, 
which are analogues of the dihedral series in $SU(2)$:
\begin{enumerate}
\item $\Delta \subset$ only the full $SU(3)$: when $n = 0 \mbox{ mod } 3$ 
	where the number of classes for $\Delta(3n^2)$
	is $(8 + \frac{1}{3}n^2)$ and for $\Delta(6n^2)$, $\frac{1}{6}(24 + 9n + n^2)$;
\item $\Delta \subset$ both the full $SU(3)$ and $SU(3)/\IZ_3$: 
	when $n \ne 0\mbox{ mod } 3$ where the number 
	of classes for $\Delta(3n^2)$ is $\frac{1}{3}(8 + n^2)$ and 
	for $\Delta(6n^2)$, $\frac{1}{6}(8 + 9n + n^2)$;
\end{enumerate}
\item Analogues of the exceptional subgroups of $SU(2)$, and indeed
like the later, there are 
two series depending on whether the $\IZ_3$-center of $SU(3)$ has been 
modded out (we recall that
the binary ${\cal T,O,I}$ are subgroups of $SU(2)$, while the ordinary 
$T,O,I$ are subgroups of the center-removed $SU(2)$, i.e., $SO(3)$,
and not the full $SU(2)$):
\begin{enumerate}
\item For $SU(3)/\IZ_3$: \\ $\Sigma_{36}, \Sigma_{60} \cong A_5$, the alternating
	symmetric-5 group, which incidentally is precisely the ordinary icosahedral group
	 $I, \Sigma_{72}, 
	\Sigma_{168} \subset S_7$, the symmetric-7 group, $\Sigma_{216} \supset 
	\Sigma_{72} \supset \Sigma_{36},$ and 
	$\Sigma_{360} \cong A_6$, the alternating symmetric-6 group;
\item For the full\footnote {In his work on Gorenstein singularities \cite {Yau}, 
	Yau points out that since the cases of 
	$\Sigma_{60 \times 3}$ and $\Sigma_{168 \times 3}$ 
	are simply direct products of the respective cases in $SU(3)/\IZ_3$ 
	with $\IZ_3$, they are usually left out by most authors. 
	The direct product simply extends the class equation
	of these groups by 3 copies and acts as an inner automorphism 
	on each conjugacy class. Therefore the character table is that of the 
	respective center-removed cases, but with the entries 
	each multiplied by the matrix $((1,1,1),(1,w,w^2),(1,w^2,w))$ 
	where $w=\exp(2 \pi i / 3)$, i.e., the full character table is the Kronecker 
	product of that of the corresponding center-removed group with that of $\IZ_3$. 
	Subsequently, the matter matrices $a_{ij}$ become the Kronecker product of 
	$a_{ij}$ for the center-removed groups with that for $\Gamma = \IZ_3$ and 
	gives no interesting new results. In light of this, 
	we shall adhere to convention and call 
	$\Sigma_{60}$ and $\Sigma_{168}$ subgroups of both $SU(3)/\IZ_3$ 
	and the full $SU(3)$ and ignore
	$\Sigma_{60 \times 3}$ and $\Sigma_{168 \times 3}$.}
	$SU(3)$: \\ $\Sigma_{36 \times 3},
	\Sigma_{60 \times 3} \cong \Sigma_{60} \times \IZ_3,
	\Sigma_{168 \times 3} \cong \Sigma_{168} \times \IZ_3,
	\Sigma_{216 \times 3},$ and $\Sigma_{360 \times 3}$.
\end{enumerate}
\end{enumerate}

Up-to-date presentations of these
groups and some character tables may be found in \cite{Yau} \cite{Fairbairn}.
The rest have been computed with \cite {Prog}.
These are included in Appendix III for reference.
As before we must narrow down our choices for ${\cal R}$. First we note
that it must be consistent with the decomposition:

\begin{equation}
\begin{array}{ccc}
SU(4) & \rightarrow & SU(3) \times U(1) \\
{\bf 4} & \rightarrow & {\bf 3}_{-1} \bigoplus {\bf 1}_{3} \\
{\bf 6} & \rightarrow & {\bf 3}_2 \bigoplus {\bf \bar{3}}_{-2} \\
\end{array}
\label{break3}
\end{equation}

This decomposition (\ref{break3}), as in the comments for (\ref{break2}),
forces us to consider only 3 dimensionals (possibly reducible) and for
the fermion case the remaining {\bf 1} must in fact be the trivial, giving
us a $\delta_{ij}$ in $a_{ij}^{\bf 4}$.

Now as far as the symmetry of $a_{ij}$ is concerned, since $SU(3)$ gives
rise to an ${\cal N} = 1$ chiral theory, the matter matrices are no longer
necessarily symmetric and we can no longer rely upon this property to guide us.
However we still have a matching condition between the bosons and the fermions.
In this ${\cal N} = 1$ chiral theory we have 2 scalars and a Weyl fermion in the 
chiral multiplet as well as a gauge field and a Weyl fermion in the vector multiplet.
If we denote the chiral and vector matrices as $C_{ij}$ and $V_{ij}$, and
recalling that there is only one adjoint field in the vector 
multiplet, then we should have:

\begin{equation}
\begin{array}{l}
a_{ij}^{\bf 4} = V_{ij} + C_{ij} = \delta_{ij} + C_{ij}\\
a_{ij}^{\bf 6} = C_{ij} + C_{ji}.\\
\end{array}
\label{break3mat}
\end{equation}

This decomposition is indeed consistent with (\ref{break3}); where the 
$\delta_{ij}$ comes from the principal {\bf 1} and the $C_{ij}$ and 
$C_{ji}$, from dual pairs of {\bf 3}; 
incidentally it also implies that the bosonic matrix should be symmetric
and that dual {\bf 3}'s should give matrices that are mutual transposes.
Finally as we have discussed in the $A_n$ case of $SU(2)$, if one is to
compose only from 1 dimensional representations, then the rows of characters for these
{\bf 1}'s must multiply identically to 1 over all conjugacy classes.
Our choices for ${\cal R}$ should thus be restricted by these general
properties.

Once again, let us analyse the groups case by case.
First the $\Sigma$ series. For the members which belong to the center-removed
$SU(2)$, as with the ordinary $T,O,I$ of $SU(2)/\IZ_2$, we expect nothing
particularly interesting (since these do not have non-trivial 3 dimensional
representations which in analogy to the non-trivial 2 dimensional irreps of
$\widehat{D_n}$ and $\widehat{E_{6,7,8}}$ should be the ones to give interesting results).
However, for completeness, we shall touch upon these groups, namely,
$\Sigma_{36,72,216,360}$. Now the {\bf 3} in (\ref{break3}) must be 
composed of {\bf 1} and {\bf 2}. The obvious choice is of course again 
the trivial one where we compose everything from only the principal {\bf 1}
giving $4 \delta_{ij}$ and $6 \delta_{ij}$ for the fermionic and bosonic 
$a_{ij}$ respectively.
We at once note that this is the only possibility for $\Sigma_{360}$, 
since its first non-trivial representation is 5 dimensional.
Hence this group is trivial for our purposes.
For $\Sigma_{36}$, the ${\bf 3}$ can come only from {\bf 1}'s for which case 
our condition that the rows must multiply to 1 implies that 
${\bf 3} = \Gamma_1 \oplus \Gamma_3 \oplus \Gamma_4$, or $\Gamma_1 \oplus \Gamma_2^2$,
both of which give uninteresting blocks, in the sense of what we have discussed
in Section 2. For $\Sigma_{72}$, we similarly
must have ${\bf 3} = \Gamma_2 \oplus \Gamma_3 \oplus \Gamma_4$ or 
${\bf 1} \oplus$ the self-dual ${\bf 2}$, both of which again give trivial
blocks. Finally for $\Sigma_{216}$, whose conjugacy classes consist 
essentially of $\IZ_3$-cycles in the 1 and 2 dimensional representations, 
the ${\bf 3}$ comes from ${\bf 1} \oplus {\bf 2}$
and the dual ${\bf 3}$, from ${\bf 1} \oplus {\bf 2^{'}}$.

For the groups belonging to the full $SU(3)$, 
namely $\Sigma_{168, 60, 36 \times 3, 216 \times 3, 360 \times 3}$, the situation is clear:
as to be expected in analogy to the $SU(2)$ case, there always exist dual 
pairs of {\bf 3} representations. The fermionic
matrix is thus obtained by tensoring the trivial representations with one member from a pair 
selected in turn out of the various pairs, i.e., ${\bf 1} \oplus {\bf 3}$; 
and indeed we have explicitly checked that 
the others (i.e., ${\bf 1} \oplus {\bf 3^{'}}$) are permutations thereof. On the
other hand, the bosonic matrix is obtained from tensoring any choice of a dual pair ${\bf 3} 
\oplus {\bf 3^{'}}$ and again we have explicitly checked that other dual pairs give rise to 
permutations. We may be tempted to construct the {\bf 3} out of the {\bf 1}'s and {\bf 2}'s
which do exist for $\Sigma_{36 \times 3, 216 \times 3}$, however we note that in 
these cases the {\bf 1} and {\bf 2} characters are all cycles of 
$\IZ_3$'s which would again give uninteresting blocks. Thus
we conclude still that for all these groups, ${\bf 4} = {\bf 1} \oplus {\bf 3}$ while
${\bf 6} = {\bf 3} \oplus$ dual ${\bf \bar{3}}$.
These choices are of course obviously in accordance with the 
decomposition (\ref{break3}) above.
Furthermore, for the $\Sigma$ groups that belong solely to the full $SU(3)$, the dual pair of 
{\bf 3}'s always gives matrices that are mutual transposes, consistent
with the requirement in (\ref{break3mat}) that the bosonic matrix be symmetric.

Moving on to the two $\Delta$ series. We note\footnote{
Though congruence in this case really means group isomorphisms, 
for our purposes since only the group characters concern us, in what follows
we might use the term loosely to mean identical character tables.}, that
for $n=1$, $\Delta_{3} \cong \IZ_3$ and $\Delta_{6} \cong d_6$ while for 
$n=2$, $\Delta_{12} \cong T := E_6$ and $\Delta_{24} \cong O := E_7$.
Again we note that for all $n > 1$ (we have already analysed the $n=1$ 
case\footnote{Of course for $\IZ_3$, we must have a different choice for ${\cal R}$,
in particular to get a good chiral model, we take the ${\bf 3} = {\bf 1'} \oplus
{\bf 1''} \oplus {\bf 1'''}$} for 
$\Gamma \subset SU(2)$), there exist
the dual {\bf 3} and ${\bf 3^{'}}$ representations as in the 
$\Sigma \subset$ full $SU(3)$ above; this is expected of course
since as noted before, all the $\Delta$ groups at least belong to the full $SU(3)$. 
Whence we again form the fermionic $a_{ij}$ from ${\bf 1} \oplus {\bf 3^{'}}$, giving
a generically nonsymmetric matrix (and hence a good chiral theory),
and the bosonic, from ${\bf 3} \oplus {\bf 3^{'}}$, giving us always a
symmetric matrix as required. We note in passing that when $n = 0 \mbox{ mod } 3$,
i.e,. when the group belongs to both the full and the center-removed
$SU(3)$, the $\Delta_{3n^2}$ matrices consist of a trivial diagonal block
and an L-shaped block. Moreover, all the $\Delta_{6n^2}$ matrices are
block decomposable. We shall discuss the significances of this observation
in the next section.
Our analysis of the discrete subgroups of $SU(3)$ is now complete; the
results are tabulated in Appendix IV.

\section{Quiver Theory? Chiral Gauge Theories?}
\begin{figure}
	\centerline{\psfig{figure=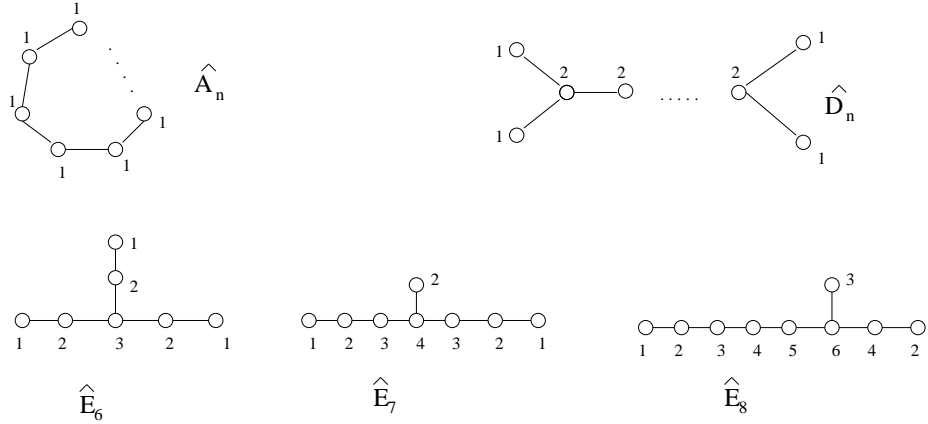,width=4.8in,angle=270}}
\caption{$\Gamma \subset$ full $SU(2)$ correspond to affine Dynkin diagrams
with the Dynkin labels $N_i$ on the nodes corresponding to the dimensions of the
irreps. In the quiver theory the nodes correspond to gauge groups and the
lines (or arrows for chiral theories), matter fields. For finite theories
each $N_i$ must be $\frac{1}{2}$ of the sum of neighbouring labels and the
gauge group is $\bigoplus\limits_i U(N_i)$. \label{su21}}
\end{figure}

\begin{figure}
	\centerline{\psfig{figure=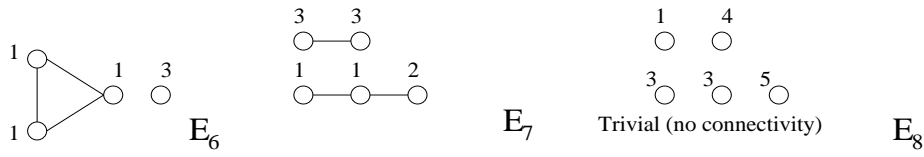,width=4.8in,angle=270}}
\caption{$\Gamma \subset SU(2)/\IZ_2$ give disconnected graphs \label{su22}}
\end{figure}

Let us digress briefly to make some mathematical observations.
We recall that in the $SU(2)$ case the matter matrices $a_{ij}$, 
due to Mckay's
theorem and Moore-Douglas quiver theories, are encoded as adjacency matrices of
affine Dynkin diagrams considered as unoriented graphs as given by Figures
\ref{su21} and \ref{su22}.


\begin{figure}
	\centerline{\psfig{figure=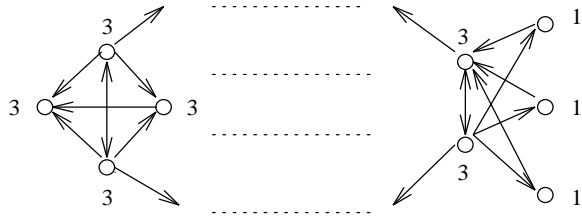,width=3.0in,angle=270}}
\caption{$\Delta_{3n^2} \subset SU(3)$ for $n \ne 0 $ mod 3. These belong
to both the full and center-removed $SU(3)$. \label{su3d}}
\end{figure}

\begin{figure}
	\centerline{\psfig{figure=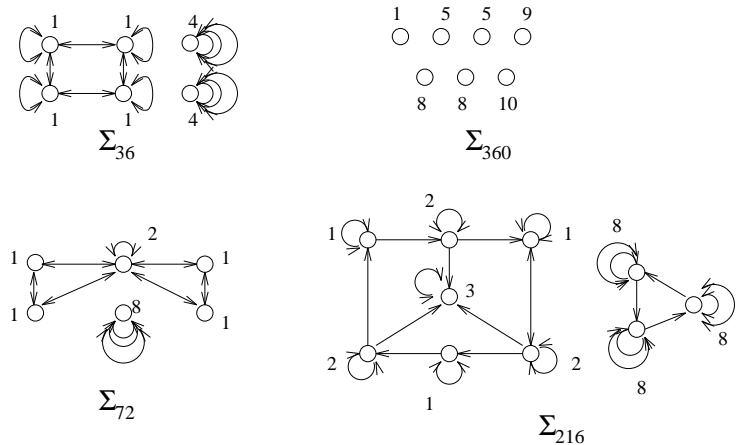,width=3.8in,angle=270}}
\caption{$\Sigma \subset SU(3)/\IZ_3$ gives unconnected graphs. \label{su3s1}}
\end{figure}

\begin{figure}
	\centerline{\psfig{figure=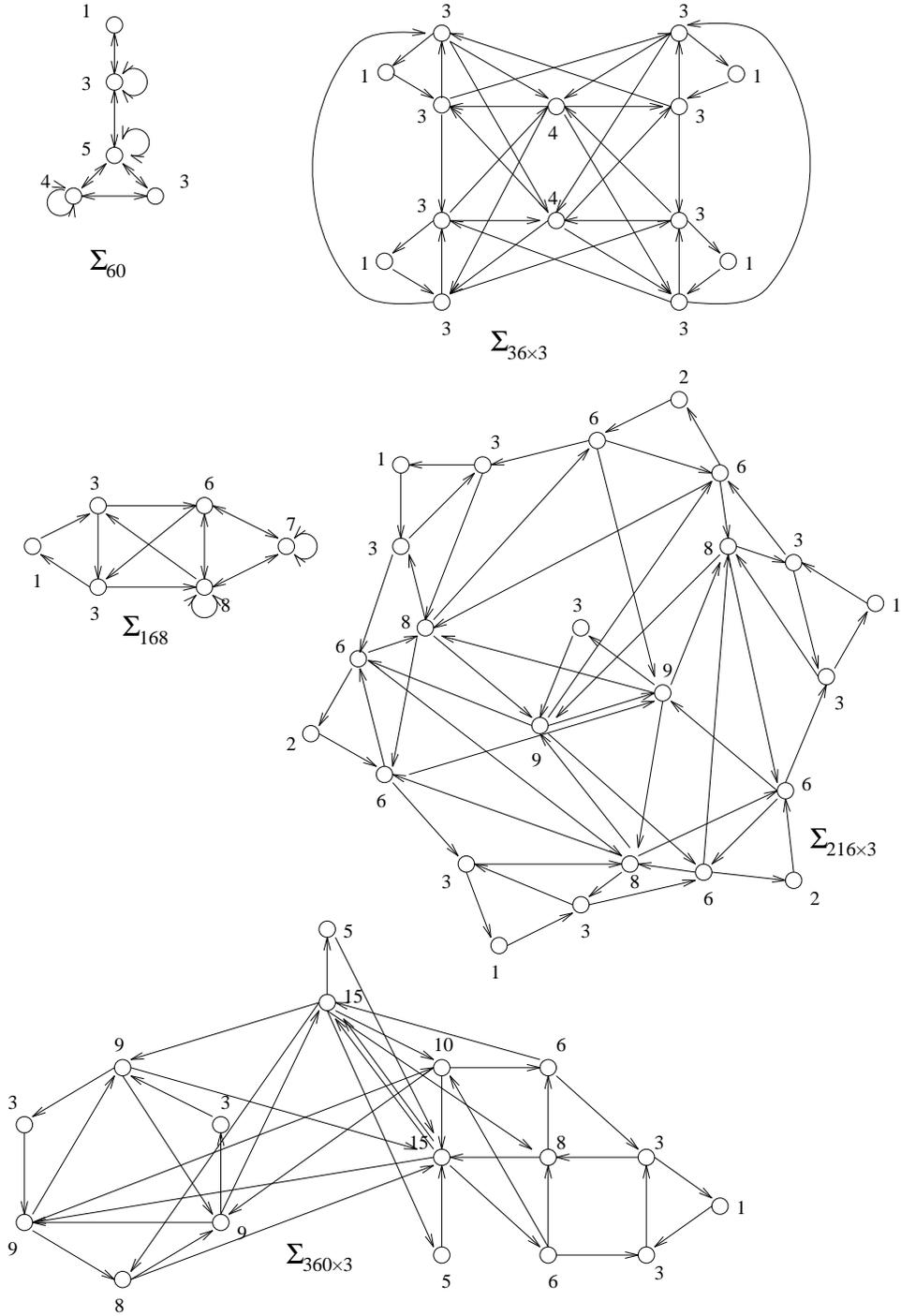,width=5.0in,angle=270}}
\caption{$\Sigma \subset$ full $SU(3)$. Only 
	$\Sigma_{36 \times 3, 216 \times 3, 360 \times 3}$ 
	belong only to the full $SU(3)$, for
	these we have the one loop $\beta$-function vanishing condition
        manifesting
	as the label of each node equaling to $\frac{1}{3}$ of that
	of the incoming and outgoing neighbours respectively.
        The matrix representation for these graphs are given in Appendix IV.
        \label{su3s2}}
\end{figure}

We are of course led to wonder, whether in analogy, the $a_{ij}$ for $SU(3)$
present themselves as adjacency matrices for quiver diagrams associated to
some {\it oriented} graph theory because the theory is chiral.
This is very much in the spirit of 
recent works on extensions of Mckay correspondences by algebraic geometers
\cite{Ito} \cite{Roan}. We here present these quiver graphs in figures
\ref{su3d} \ref{su3s1} and \ref{su3s2}, hoping 
that it may be of academic interest.


Indeed we note that for the center-removed case, as with $SU(2)$, we get disconnected
(or trivial) graphs; this of course is the manifestation of the fact that
there are no non-trivial {\bf 3} representations for these groups (just
as there are no non-trivial {\bf 2}'s of $\Gamma \subset SU(2) / \IZ_2$).
On the other hand for $\Gamma \subset$ full $SU(3)$, we do get interesting connected and
oriented graphs, composed of various directed triangular cycles.

Do we recognise these graphs? The answer is sort of yes and the right
place to look for turns out to be in conformal field theory. 
In the work on general modular invariants in the WZW model for
$\widehat{su(n)_k}$ (which is equivalent to the study of the modular properties
of the characters for affine Lie algebras), an $ADE$ classification was noted
for $n=2$ \cite{CFT} \cite{Gepner} \cite{Bernard}; this should somewhat be expected due to our 
earlier discussion on Gabriel's Theorem. For $n=3$, work has been done to
extract coefficients in the fusion rules and to treat them as entries of
adjacency matrices; this fundamentally is analogous to what we have done since
fusion rules are an affine version of finite group composition coefficients.
So-called generalised Dynkin diagrams have been constructed for 
$\widehat{su(3)}$ in analogy to the 5 simply-laced types corresponding to
$SU(2)$, they are: ${\cal A}_n, {\cal D}_{3n}, {\cal E}_5, {\cal E}_9$ and
${\cal E}_{21}$ where the subscripts denote the level in the representation
of the affine algebra \cite{CFT} \cite {DiFrancesco} \cite{Gannon}. 
We note a striking resemblance between these graphs
(they are some form of a dual and we hope to rigorise this similarity in
future work) with our quiver graphs: the ${\cal E}_5, {\cal E}_9$ and
${\cal E}_{21}$ correspond to $\Sigma_{216 \times 3}$, 
$\Sigma_{360 \times 3}$, and $\Sigma_{36 \times 3}$ respectively. Incidentally
these $\Sigma$ groups are the {\it only} ones that belong solely to the
full and not the center-removed $SU(3)$. The ${\cal D}_{3n}$ corresponds to
$\Delta_{3n^2}$ for $n \ne 0\mbox{ mod }3$, which are the non-trivial ones as
observed in the previous section and which again are those that belong
solely to the full $SU(3)$. The $\Delta_{6n^2}$ series, as noted above,
gave non-connected graphs, and hence do not have a correspondent. Finally
the ${\cal A}_n$, whose graph has complete $\IZ_3$ symmetry must come from
the Abelian subgroup of $SU(3)$, i.e., the $\widehat{A_n}$ case of $SU(2)$
but with ${\cal R} = {\bf 3}$ and not {\bf 2}. This beautiful relationship
prompts us to make the following conjecture upon which we may labour in
the near future:\\

{\large {\bf Conjecture:}} There exists a McKay-type correspondence
between Gorenstein singularities and the characters
of integrable representations of affine algebras $\widehat{su(n)}$
(and hence the modular invariants of the WZW model).\\

A physical connection between $\widehat{SU(2)}$ modular invariants and
quiver theories with 8 supercharges has been pointed out \cite{SD}. 
We remark that our conjecture is in the same spirit and a hint may come from
string theory. If we consider a D1 string on our orbifold, then this is
just our configuration of D3 branes after two T-dualities. In the strong
coupling limit, this is just an F1 string in such a background which
amounts to a non-linear sigma model and therefore some (super) conformal
field theory whose partition function gives rise to the modular invariants.
Moreover, connections between such modular forms and Fermat varieties
have also been pointed out \cite{Fermat}, this opens yet another door for
us and many elegant intricacies arise.

Enough digression on mathematics; let us return to physics. We would like
to conclude by giving a reference catalogue of chiral 
theories obtainable from
$SU(3)$ orbifolds. Indeed, though some of the matrices may not be terribly
interesting graph-theoretically, the non-symmetry of $a_{ij}^{\bf 4}$
is still an indication of a good chiral theory.

For the original $U(n)$ theory it is conventional to take a canonical
decomposition \cite{LNV}
as $n=N|\Gamma|$ \cite{LNV}, whence the (orbifolded) gauge group 
must be $\bigotimes\limits_{i}SU(N_i)$ as discussed in Section 3, such that 
$N|\Gamma|=n=\sum\limits_i N_i |{\bf r}_i|$. By an elementary theorem on
finite characters: $|\Gamma|=\sum \limits_i |{\bf r}_i|^2$, we see that the
solution is $N_i=N|{\bf r}_i|$. This thus immediately gives the form of 
the gauge group. Incidentally for $SU(2)$, the McKay
correspondence gives more information, it dictates that the dimensions
of the irreps of $\Gamma$ are actually the Dynkin labels for the diagrams.
This is why we have labeled the nodes in the graphs above. Similarly for
$SU(3)$, we have done so as well; these should be some form of generalised
Dynkin labels.

Now for the promised catalogue, we shall list below all the 
chiral theories obtainable from orbifolds of $\Gamma \subset SU(3)$ 
($\IZ_3$ center-removed or not). This is
done so by observing the graphs, connected or not, that contain unidirectional
arrows. For completeness, we also include the subgroups of $SU(2)$, which
are of course also in $SU(3)$, and which do give non-symmetric matter matrices
(which we eliminated in the ${\cal N}=2$ case) if we judiciously choose the
{\bf 3} from their representations. We use the short hand 
$(n_1^{k_1},n_2^{k_2},...,n_i^{k_i})$ to denote the gauge group
$\bigoplus\limits^{k_1} SU(n_1)...\bigoplus\limits^{k_i} SU(n_i)$. Analogous to
the discussion in Section 3, the conformality condition to one loop order
in this ${\cal N} = 1$
case, viz., $N_f = 3 N_c$ translates to the requirement that the label of
each node must be $\frac13$ of the sum of incoming and the sum of 
outgoing neighbours individually.  (Incidentally, the gauge anomaly cancelation 
condition has been pointed out as well \cite{LR}. In our language it demands the
restriction that $N_j a_{ij} = \bar{N}_j a_{ji}$.)
In the following table, the * shall
denote those groups for which this node condition is satisfied. We see that
many of these models contain the group $SU(3) \times SU(2) \times U(1)$ and hope
that some choice of orbifolds may thereby contain the Standard Model.
\[
\begin{array}{|c|c|}
\hline
\Gamma \subset SU(3)				&\mbox{Gauge Group} \\ \hline
\widehat{A_n} \cong \IZ_{n+1}			&(1^{n+1}) \\
\IZ_k \times \IZ_{k'}				&(1^{k k'}) *\\
\widehat{D_n}					&(1^4,2^{n-3}) \\
\widehat{E_6} \cong {\cal T}			&(1^3,2^3,3) \\
\widehat{E_7} \cong {\cal O}			&(1^2,2^2,3^2,4) \\
\widehat{E_8} \cong {\cal I}			&(1,2^2,3^2,4^2,5,6) \\
E_6 \cong T					&(1^3,3) \\
E_7 \cong O					&(1^2,2,3^2) \\
E_8 \cong I					&(1,3^2,4,5) \\
\Delta_{3n^2}	(n=0\mbox{ mod }3)	  	&(1^9,3^{\frac{n^2}{3}-1}) *\\
\Delta_{3n^2}	(n\ne0\mbox{ mod }3)		&(1^3,3^{\frac{n^2-1}{3}}) *\\
\Delta_{6n^2}	(n\ne0\mbox{ mod }3) 		&(1^2,2,3^{2(n-1)},6^{\frac{n^2-3n+2}{6}}) *\\
\Sigma_{168}					&(1,3^2,6,7,8) *\\
\Sigma_{216}					&(1^3,2^3,3,8^3) \\
\Sigma_{36\times 3}				&(1^4,3^8,4^2) *\\
\Sigma_{216\times 3}				&(1^3,2^3,3^7,6^6,8^3,9^2) *\\
\Sigma_{360\times 3}				&(1,3^4,5^2,6^2,8^2,9^3,10,15^2) *\\
\hline
\end{array}
\]

\section{Concluding Remarks}
By studying gauge theories constructed from orbifolding of an ${\cal N}=4$
$U(n)$ super-Yang-Mills theory in 4 dimensions, we have touched upon
many issues. We have presented the explicit matter content and gauge
group that result from such a procedure, for the cases of $SU(2)$ and
$SU(3)$. In the
first we have shown how our calculations agree with current quiver 
constructions and
in the second we have constructed possible candidates for chiral theories.
Furthermore we have noted beautiful graph-theoretic interpretations of these
results: in the $SU(2)$ we have used Gabriel's theorem to partially explain
the $ADE$ outcome and in the $SU(3)$ we have noted connections with 
generalised Dynkin diagrams and have conjectured the
existence of a McKay-type correspondence between these orbifold theories
and modular invariants of WZW conformal models.

Much work of course remains. In addition to proving this conjecture, we also
have numerous questions in physics. What about $SU(4)$, the full group?
These would give interesting non-supersymmetric theories. How do we
construct the brane box version of these theories? Roan has shown how
the Euler character of these orbifolds correspond to the class numbers
\cite{Roan}; we know the blow-up of these singularities correspond to
marginal operators. Can we extract the marginal couplings and thus the
duality group this way? We shall hope to address these problems in forth-coming
work. Perhaps after all, string orbifolds, gauge theories, modular
invariants of conformal field theories as well as Gorenstein singularities
and representations of affine Lie algebras, are all manifestations of
a fundamental truism.

\section*{Acknowledgements}
{\it Ad Catharinae Sanctae Alexandriae et Ad Majorem Dei Gloriam...\\}
We like to extend our sincere gratitude to Prof.\ M.\ Artin of the Dept.\ of
Mathematics, M.I.T., for his suggestions and especially to Prof.\ W.\ Klink of the
University of Iowa for his kind comments and input on the characters of the $SU(3)$
subgroups. 
YH would also like to thank his collegues K.\ Bering, B.\ Feng, J.\ S.\ Song and
M.\ B.\ Spradlin for valuable discussions, M.\ Serna and L.\ Williamson
for charming diversions, as well as the NSF for her gracious 
patronage.\\
This work was supported by the U.S.\ Department of 
Energy under contract \#DE-FC02-94ER40818.

\pagebreak
\setlength{\textheight}{8.75in}          
\topmargin -0.5in

\section*{Appendix I, Character Tables for the Discrete Subgroups of $SU(2)$}
Henceforth we shall use $\Gamma_i$ to index the representations and the numbers
in the first row of the character tables shall refer to the order of each conjugacy 
class, or what we called $r_\gamma$.

\noindent
{\tiny
{\large $\widehat{A}_{n}=$ Cyclic ${\IZ}_{n+1}$ \\}
\[

\]

\noindent
{\normalsize
\\
For reference, next to each of the binary groups,
we shall also include the character table of the corresponding ordinary cases,
which are in $SU(2)/\IZ_2$.\\
}

\noindent
{\large $\widehat{D}_{n}$ = Binary Dihedral}
\[
\begin{array}{|c|c|c|c|c|c|c|c|c|}
\hline
	   & 1 & 1  & 2 & \cdots  & 2 & n & n \\ \hline
\Gamma_{1} & 1 & 1  & 1 & \cdots  & 1 & 1 & 1 \\ \hline
\Gamma_{2} & 1 & 1  & 1 & \cdots  & 1 & -1 & -1 \\ \hline
\Gamma_{3} & 1 & (-1)^n & (-1)^1  & \cdots  & (-1)^{n-1} & i^n & -i^n \\ \hline
\Gamma_{4} & 1 & (-1)^n & (-1)^1  & \cdots  & (-1)^{n-1} & -i^n  & i^n \\ \hline
\Gamma_{5} & 2 & (-2)^1 & 2\cos \frac{\pi}{n}  & \cdots  & 
	2\cos\frac{\pi(n-1)}{n}  & 0 & 0 \\ \hline
	\vdots &\vdots  &\vdots  &\cdots  &\vdots  &\vdots  &\vdots  &\vdots \\ \hline
\Gamma_{n+1} & 2 & (-2)^{n-1} & 2\cos\frac{\pi(n-1)}{n}  &
	\cdots  & 2\cos\frac{\pi (n-1)^{2}}{n}  & 0 & 0 \\ \hline
\end{array}%
\]
}

\section*{Appendix II, Matter Content for ${\cal N}=2$ SUSY Gauge
Theory ($\Gamma \subset SU(2)$)}
Only the fermionic matrices are presented here; as can be seen from the
decomposition, twice the fermion $a_{ij}$ subtracted by $2\delta_{ij}$ 
should give the bosonic counterparts, which follows from supersymmetry.
In the ensuing, {\bf 1} shall denote the (trivial) principal representation,
${\bf 1^{'}}$ and ${\bf 1^{''}}$, dual (conjugate) pairs of 1 dimensional
representations.

{\tiny

\[
%
\]
}

\setlength{\textwidth}{8.0in}           
\oddsidemargin -0.5in
\evensidemargin -0.5in

\section*{Appendix III, Classification of Discrete Subgroups of $SU(3)$}
\subsection* {Type I: The $\Sigma$ Series}
These are the analogues of the $SU(2)$ crystallographic groups and their 
double covers, i.e., the E series. We have:
\[
\begin{array}{ccc}
\Sigma_{36}, \Sigma_{72}, \Sigma_{216}, \Sigma_{60}, \Sigma_{168},
\Sigma_{360} & \subset & SU(3) / (\IZ_3 $ center$) \\
\Sigma_{36 \times 3}, \Sigma_{72 \times 3}, \Sigma_{216 \times 3}, 
\Sigma_{60 \times 3}, \Sigma_{168 \times 3}, \Sigma_{360 \times 3} & \subset & SU(3)
\end{array}
\]

\subsubsection*{Type Ia: $\Sigma \subset SU(3) / \IZ_3 $}
The character tables for the center-removed case have been given by 
\cite{Fairbairn}.
{\tiny

\[
%
\]
}

\subsubsection*{Type Ib: $\Sigma \subset $ full $SU(3)$}
The character tables are computed, using \cite {Prog}, from the generators
presented in \cite {Yau}. In what follows, we define 
$e_n = \exp{\frac{2 \pi i}{n}}$.

{\tiny
\[
%
\]
}

\subsubsection*{Type Ic: $\Sigma \subset$ both $SU(3)$ and $SU(3)/\IZ_3$}

{\tiny
\[
%
\]
}

\subsection*{Type II: The $\Delta$ series}
These are the analogues of the dihedral subgroups of $SU(2)$ (i.e., the
D series).\\
{\large $\Delta_{3n^2}$}
\[
\begin{array}{cccc}
		  	& $Number of classes$	& $Subgroup of$ & $Some Irreps$ \\ \hline
n = 0 $ mod $ 3 	& 8 + \frac{1}{3}n^2   	& $Full $ SU(3)  
			& 9  $ {\bf 1}'s$ \frac13n^2-1 $ {\bf 3}'s$ \\
n \ne 0 $ mod $ 3 	& \frac{1}{3}(8 + n^2)   & $Full $ SU(3) $ and $ SU(3)/\IZ_3  
			& 3 $ {\bf 1}'s$, \frac{1}{3}(n^2 -1) $ {\bf 3}'s$	\\
\end{array}%
\]

\section*{Appendix IV, Matter content for $\Gamma \subset SU(3)$}
Note here that since the ${\cal N} = 1$ theory is chiral, the fermion
matter matrix need not be symmetric. A graphic representation for some of these
theories appear in figures 3, 4 and 5.

{\tiny

\[
%
\]
}

\end{document}